\documentclass[12pt]{article}

\usepackage{epsfig}
\usepackage{latexsym}

\begin{document}

\begin{titlepage}

\baselineskip 24pt

\begin{center}

{\Large {\bf Circumstantial Evidence for Rotating Mass Matrix from Fermion 
   Mass and Mixing Data}}

\vspace{.5cm}

\baselineskip 14pt

{\large Jos\'e BORDES}\\
jose.m.bordes\,@\,uv.es\\
{\it Departament Fisica Teorica, Universitat de Valencia,\\
  calle Dr. Moliner 50, E-46100 Burjassot (Valencia), Spain}\\
\vspace*{.4cm}
{\large CHAN Hong-Mo}\\
chanhm\,@\,v2.rl.ac.uk \\
{\it Rutherford Appleton Laboratory,\\
  Chilton, Didcot, Oxon, OX11 0QX, United Kingdom\\
and\\
NAPL, Department of Physics,\\ University of Oxford, Oxford, United 
Kingdom}\\
\vspace*{.4cm}
{\large TSOU Sheung Tsun}\\
tsou\,@\,maths.ox.ac.uk\\
{\it Mathematical Institute, University of Oxford,\\
  24-29 St. Giles', Oxford, OX1 3LB, United Kingdom}

\end{center}

\vspace{.3cm}

\begin{abstract}

It is shown that existing data on the mixing between up and down fermion
states and on the hierarchical mass ratios between fermion generations, 
as far as can be so analysed at present, are all consistent with the two 
phenomena being both consequences of a mass matrix rotating in generation
space with changing energy scale.  As a result, the rotating mass matrix
can be traced over some 14 orders of magnitude in energy from the mass
scale of the $t$-quark at 175 GeV to below that of the atmospheric neutrino 
at 0.05 eV. 

\end{abstract}

\end{titlepage}

\clearpage

\baselineskip 14pt

\section{Introduction}

Along with the mystery of why there should be in nature 3, and apparently 
only 3, generations of fermions, the fact that their masses should be
hierarchical and that they should mix, as embodied for quarks in the CKM 
matrix \cite{CKM} and exhibited for leptons in neutrino oscillations, has 
remained one of the great puzzles of particle physics.  In the context of 
what we call the Dualized Standard Model (DSM) \cite{dualgen} which is an 
explicit attempt to solve this generation puzzle, we have suggested that 
the mass hierarchy and the mixing phenomenon can both result from a mass 
matrix which changes its orientation in generation space (rotates) with 
changing energy scale and have obtained rather good agreement with experiment 
based on this hypothesis.  This previous treatment, however, does not
fully cover the following two aspects.  Firstly it depends  
on the details of the DSM mechanism driving the rotation, which details
may not be strictly necessary for deriving the said result, and secondly, it 
does not clearly reveal the degree of significance of the claimed agreement 
with experiment nor the amount of direct empirical support, if any, for 
mass matrix rotation.  For this reason, our purpose in this paper is to 
leave aside for the moment the driving mechanism (whether DSM or otherwise) 
but put instead the rotating mass matrix directly to the test by going 
straight to the experimental data and seek evidence there for the rotation 
hypothesis.

We shall show that by inputting all the available mass and mixing data 
on both quarks and leptons, assuming only that these all arise from mass 
matrix rotation in the manner to be explained below, one can trace the 
implied rotation over a scale range of some 14 orders of magnitude.  The 
result is seen to be fully consistent with the fermion states all
lying on a single smooth 
rotation curve linking the $t$ quark at 175 GeV through all the intermediate 
fermion states down to the second heaviest neutrino $\nu_2$ at less than 
$10^{-2}\ {\rm eV}$.  This constitutes positive evidence for mass matrix 
rotation, although as yet only circumstantial, which is independent of the 
theoretical mechanism responsible for driving it, whether DSM or otherwise.  
It will be seen, nevertheless, that the shape of the empirical rotation 
curve so traced is indicative of having the two rotational fixed points 
at infinite and zero energy scales predicted by and, as far as we know, 
special to the DSM scheme. 

We begin by re-examining the reasoning behind the ansatz that both mass
hierarchy and mixing can arise from a rotating mass matrix, for although 
it has been stated here and there
and used already in several earlier papers, we cannot 
assume that it is of general knowledge.  The first point to note is that 
once the mass matrix rotates with changing scale, then even such familiar 
concepts as particle masses, state vectors and mixing of flavour states, 
will have to be revised or refined, as we shall now explain. 

Let us start with a fermion mass matrix traditionally defined by a term 
in the action of the form:
\begin{equation}
\bar{\psi}^0_L m \psi^0_R + {\rm h.c.},
\label{mtrad}
\end{equation}
where $\psi^0_L$ and $\psi^0_R$ represent respectively the left- and
right-handed fermion field, each being a vector in 3-dimensional flavour
space, here given in the weak gauge basis, and $m$ is a $3 \times 3$
(complex) matrix.  The matrix $m$ can always be diagonalized thus:
\begin{equation}
U_L^\dagger m U_R = {\rm diag}\{m_1, m_2, m_3\}
\label{diagmtrad}
\end{equation}
with $U_L, U_R$ unitary and $m_i$ real.  Thus in terms of the fields:
\begin{equation}
\psi_L = U_L^\dagger \psi^0_L; \ \ \ \psi_R = U_R^\dagger \psi^0_R,
\label{mtradev}
\end{equation}
the term (\ref{mtrad}) in the action takes on the diagonal form:
\begin{equation}
\bar{\psi}_L {\rm diag}\{m_1, m_2, m_3\} \psi_R.
\label{mtraddiag}
\end{equation}
When the mass matrix $m$ is constant in orientation with respect to scale 
change, i.e.\ in the language of this paper, when the mass matrix does not 
rotate, which is the simple case usually considered, then the particle 
masses of the 3 flavour states are just given by the diagonal values $m_i$.  
The above apply to both up and down quarks in the case of quarks, and to 
both charged leptons and neutrinos in the case of leptons.  Hence, from 
the mass matrix, one obtains for the up and down states each a diagonalizing 
matrix $U_L$ which we can denote respectively as $U_L$ and $U'_L$.  Again, 
in the simple case when the mass matrices do not rotate, then the mixing 
matrix between up and down states (i.e CKM \cite{CKM} for quarks and MNS 
\cite{MNS} for leptons) is given just as \cite{Jarlskog}:
\begin{equation}
V = U_L {U'}^\dagger_L.
\label{mixingmJ}
\end{equation}

For the discussion in this paper, it is more convenient to work with an
equivalent form of the mass matrix adopted by Weinberg in \cite{Weinberg}.
Since the right-handed fermion fields are flavour singlets, they can be
arbitrarily relabelled without changing any of the physics. (Witness the 
fact that the mixing matrices between up and down states depend only on
$U_L$ not on $U_R$.)  Hence, by an appropriate relabelling of right-handed 
fields, explicitly by defining new right-handed fields:
\begin{equation}
{\psi'}^0_R = U_L U_R^\dagger \psi^0_R,
\label{newpsir}
\end{equation}
one obtains (\ref{mtrad}) in a form in which the the mass matrix becomes
Hermitian:
\begin{equation}
\bar{\psi} m_W \frac{1}{2}(1 + \gamma_5) \psi
   + \bar{\psi} m_W \frac{1}{2}(1 - \gamma_5) \psi = \bar{\psi} m_W \psi,
\label{mtradW}
\end{equation}
with
\begin{equation}
m_W = m U_R U_L^\dagger.
\label{mW}
\end{equation}
This is convenient because in the simple case when the mass matrix does 
not rotate, the particle masses are now just the real eigenvalues of the
Hermitian matrix $m_W$, as can readily be checked with (\ref{diagmtrad}).
Furthermore, the mixing matrix between up and down states become just
\begin{equation}
V_{ij} = \langle {\bf v}_i|{\bf v'}_j \rangle,
\label{ckmdot}
\end{equation}
with $|{\bf v}_i \rangle$ being the eigenvector of $m_W$ corresponding to the 
eigenvalue $m_i$, and the prime denoting down-type fermions.
The scalar product 
$\langle {\bf v}_i|{\bf v'}_j \rangle$ is of 
course an invariant independent of the frame in which these vectors 
are expressed.  Thus, in terms of this Weinberg form of 
the mass matrix $m_W$ (which in this paper will be used exclusively and 
from which the subscript $W$ for convenience will henceforth be omitted), 
the mass values and state vectors of flavour states, as well as the mixing 
matrix between up and down fermions, are all easily defined in the case 
of no rotation in the mass matrix.

Consider now what happens in the case when the mass matrix rotates with
changing scale as examined in this paper.  Both its eigenvalues and
their corresponding eigenvectors now change with the scale so that the
previous definition of these as respectively the masses and state vectors
of flavour states is no longer sufficiently precise, for it will have to
be specified at which scale(s) the eigenvalues and eigenvectors are to be 
evaluated.

In the simple case of a single generation, i.e. when the the mass matrix
is just a number, one is used to defining the particle mass as the running
mass taken at the scale equal to the mass value itself, i.e. at that $\mu$
when $\mu = m(\mu)$.  One might be tempted therefore to suggest for the
multi-generation case that one defines the mass $m_i$ and the state vector 
${\bf v}_i$ of the state $i$, as respectively just the $i$th eigenvalue and 
eigenvector of the matrix $m$ taken at the scale $\mu_i = m_i(\mu_i)$, with
$m_i(\mu)$ being the scale-dependent $i$th eigenvalue of the matrix $m$.
However, such a definition will not do, because it would mean that the state 
vectors for the different generations $i$ will be defined as eigenvectors 
of the matrix $m$ at different scales.  Although the 
eigenvectors $i$ for different eigenvalues $i$ are orthogonal, 
$m$ being Hermition, when taken 
all at the same scale, they need not be mutually orthogonal when taken each 
at a different scale.  But the state vectors for different flavour states
ought to be orthogonal to one another if they are to be independent quantum 
states.  Otherwise, it would mean physically that the flavour states (such 
as the charged leptons $e, \mu$ and $\tau$, for example) would have nonzero 
components in each other and be thus freely convertible into one another, 
or that the mixing matrices (CKM for quarks and MNS for leptons) would no
longer be unitary, which would of course be unphysical.

How then should the mass values and state vectors of flavour states be
defined in the scenario when the mass matrix rotates?  To see how this 
question may be resolved, let us examine it anew with first the $U$-type 
quarks as example.  The $3 \times 3$ mass matrix $m$ has 3 eigenvalues 
with the highest value $m_1$ corresponding to the eigenvector ${\bf v}_1$, 
both depending on scale $\mu$.  Starting from a high scale and proceeding 
lower, one reaches at some stage $\mu_1 = m_1(\mu_1)$, i.e. when the scale 
equals the highest eigenvalue $m_1$.  One can then naturally define this 
value $m_1(\mu_1)$ as the $t$ quark mass $m_t$ and the corresponding 
eigenvector ${\bf v}_1(\mu_1)$ as the $t$ state vector ${\bf v}_t$.  Next, 
how should one define the mass $m_c$ and the state vector ${\bf v}_c$?  We 
have already seen above that they cannot be defined as respectively the
second highest eigenvalue $m_2$ of the $3 \times 3$ mass matrix $m$ and 
its corresponding eigenvector at the scale $\mu_2 = m_2(\mu_2)$, because 
this vector is in general not orthogonal to the state vector ${\bf v}_t$
which the state vector ${\bf v}_c$ ought to be.  
It is not difficult, however, to 
see what is amiss.  At scales below the $t$ mass, i.e. when $\mu < m_t$, 
$t$ would no longer exist as a physical state, so that what functions 
there as the fermion mass matrix is not the $3 \times 3$ matrix $m$ but 
only the $2 \times 2$ submatrix, say $\hat{m}$, of $m$ in the subspace 
orthogonal to ${\bf v}_t$.  Hence, for consistency, one should define  
$m_c$ as the highest eigenvalue $\hat{m}_2$ of the matrix $\hat{m}$ and 
the state vector ${\bf v}_c$ as the corresponding eigenvector, both at 
the scale $\hat{\mu}_2 = \hat{m}_2(\hat{\mu}_2)$.  The state vector of
$c$ so obtained is of course automatically orthogonal to ${\bf v}_t$ as
it should be.  Repeating the argument, one defines further the mass $m_u$ 
and state vector ${\bf v}_u$ respectively as the ``eigenvalue'' and 
``eigenvector'' of $\hat{\hat{m}}$ at the scale $\hat{\hat{\mu}}_3 = 
\hat{\hat{m}}_3(\hat{\hat{\mu}}_3)$, with $\hat{\hat{m}}$ being the 
$1 \times 1$ submatrix of $m$ in the subspace orthogonal to both 
${\bf v}_t$ and ${\bf v}_c$.  Proceeding in this way, all masses and state 
vectors are defined at their own proper mass scale and the state vectors 
are mutually orthogonal as they should be.  Besides, though stated above 
only for 3, the definition can be extended to any number of fermion 
generations, should there be physical incentive for doing so.

We note that a refined definition of masses and state vectors for flavour 
states is a question which has to be adressed in principle when the mass 
matrix rotates with changing scale, whatever the speed of the rotation.
Hence, it ought to figure even in the conventional formulation of the 
Standard Model where the mass matrix is bound to rotate when there is 
nontrivial mixing between up and down states \cite{Ramon,physcons}, 
although the rotation there is rather slow and its effects are for most 
practical applications negligible.  The above solution to the problem, 
though first made in the context of the DSM scheme \cite{physcons}, is 
seen actually to apply to any rotating mass matrix, and it is, as far as 
we know, the only solution to the question yet given in the literature.  
So this is the definition we shall employ in what follows to analyse 
existing mass and mixing data in terms of a rotating mass matrix.  The 
success or otherwise of the analysis could thus itself be regarded as a 
check on the validity of the above solution.

Having now made clear our general procedure for defining masses and state 
vectors, let us return to the problem at hand, namely that of testing 
empirically the hypothesis that both mass hierarchy and mixing arise as
consequences of the rotation of the mass matrix.  First, by `mixing arising 
from rotation', we mean that the mass matrices of up and down fermions 
taken at the same scale are 
aligned in orientation at all scales, only differing by a normalization 
factor, and it is the rotation alone which is giving rise to the mixing.  
Secondly, by `mass mass hierarchy arising from rotation' as well, we mean 
that the mass matrix has at all scales only one massive eigenstate, with 
the masses of the lower generations appearing only by virtue of the
rotation via what we called the ``leakage'' mechanism to be explained 
below.  We have thus for both up and down fermions a mass matrix of the 
form:
\begin{equation}
m = m_T |{\bf r}\rangle \langle {\bf r}|,
\label{mfact}
\end{equation}
where $m_T$ is the single nonvanishing eigenvalue of $m$ and $|{\bf r}
\rangle$
its corresponding (normalized)
eigenvector.  Or explicitly for 3 generations, we have:
\begin{equation}
m = m_T \left( \begin{array}{ccc} \xi \\ \eta \\ \zeta \end{array} 
\right) (\xi,\eta,\zeta),
\label{mfactex}
\end{equation}
where only $m_T$ depends on the fermion species.  Thus the whole content 
of the rotating mass matrix is now encapsulated in 
${\bf r} = (\xi, \eta, \zeta)$, a
vector rotating in 3-dimensional generation space with changing scale $\mu$.

\begin{figure} [ht]
\centering
\input{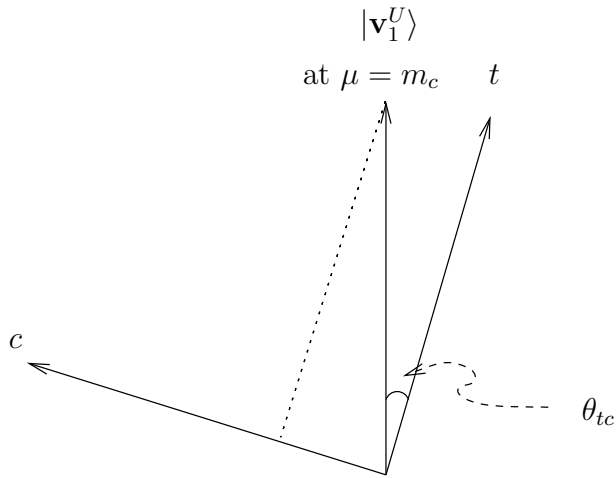}
\caption{Masses for lower generation fermions from a rotating mass matrix
via the ``leakage'' mechanism.}  
\label{planeleak}
\end{figure}

\begin{figure} [ht]
\centering
\input{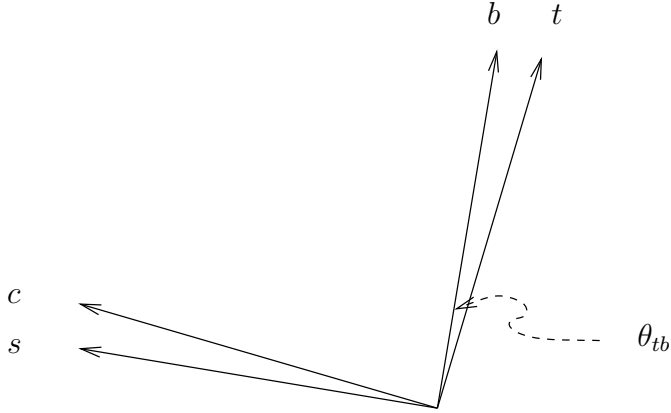}
\caption{Mixing between up and down fermions from a rotating mass matrix.}  
\label{planemix}
\end{figure}

That such a simple form of the mass matrix can nevertheless give rise to 
nontrivial mixing and mass hierarchy is most easily seen in a simplified 
situation when there are only 2 generations instead of 3.  When applied 
to this simple case, the procedure detailed above for defining masses 
and state vectors of flavour states gives the state vector ${\bf v}_t$ 
of $t$ as the single massive eigenstate ${\bf r}$ of the $U$-quark mass 
matrix at the scale $\mu = m_t$, and the vector ${\bf v}_c$ as a vector 
orthogonal to ${\bf v}_t$, as depicted in Figure \ref{planeleak}.  Thus
${\bf v}_c$ has by (\ref{mfact}) a zero eigenvalue for $m$ at the scale 
$\mu = m_t$.  But according to our previous conclusion, this should not 
be interpreted to mean that $c$ has a zero mass.  The $c$ mass $m_c$ is 
given instead as the eigenvalue of $\hat{m}$ at scale $\mu = m_c$, which 
in this simplified 2 generation case is just the expectation value of $m$ 
in the state ${\bf v}_c$.  At the scale $\mu = m_c$, however, the vector 
${\bf r}$ is already rotated to a direction different from that of 
${\bf v}_t$, as illustrated in Figure \ref{planeleak}, and has acquired a 
component equal to $\sin \theta_{tc}$ in the direction of ${\bf v}_c$,
with $\theta_{tc}$ being the rotation angle between the scales $\mu = m_t$
and $\mu = m_c$.  Hence, according to the above definition, one obtains 
``by leakage'' a nonzero mass for $c$ given by the expectation value of 
(\ref{mfact}) with respect to ${\bf v}_c$, namely:
\begin{equation}
m_c = m_t \sin^2 \theta_{tc}.
\label{planelk}
\end{equation}
Similarly, although the mass matrices of the $U$ and $D$ quarks according 
to (\ref{mfact}) are aligned in orientation at all scales, one sees from 
Figure \ref{planemix} that by virtue of the rotation of the vector ${\bf r}$ 
from the scale $\mu = m_t$ to the scale $\mu = m_b$ where the state vectors 
${\bf v}_t$ and ${\bf v}_b$ are respectively defined, the two state vectors 
will not point in the same direction.  One has thus from (\ref{ckmdot})
simply by virtue of the rotation a nonzero mixing between the $t$ and $b$ 
states with the CKM matrix element:
\begin{equation}
V_{tb} = {\bf v}_t.{\bf v}_b = \cos \theta_{tb} \neq 1, 
\label{planemx}
\end{equation}
where $\theta_{tb}$ is the rotation angle between the two scales.

One sees already from these examples in the simplified scenario of only 2
generations that both lower generation masses and nontrivial mixing will 
automatically be obtained from a rotating mass matrix even if one starts 
with neither.  One sees also that the same conclusion can be drawn in the 
3 generation case so long as the masses and state vectors of flavour states 
are defined for a rotating mass matrix as detailed above.  The remaining 
question to ask is then whether the actual masses and mixing of fermion states
as observed in experiment can actually be explained in this simple way from 
a rotating mass matrix, and this is the question we set out to answer in 
this paper.  This question can be adressed empirically since, as we shall 
show, many of the relevant quantities have already been measured and need 
only to be arranged and interpreted appropriately for the present purpose 
according to the definitions detailed above.

\section{Analysis in the planar approximation}    

Let us first perform the analysis in the simplified situation with only 
the 2 heaviest generations in each fermion species, which scenario makes
the analysis much more transparent since the problem then becomes planar 
and there is only one real rotation angle to consider \cite{CKM,Jarlskog}.
This simplification will be shown later to approximate already rather well 
the actual 3-generation situation in the scale region above roughly the 
$\mu$ lepton
mass.  We have then the pictures shown in Figures \ref{planeleak} and 
\ref{planemix} for obtaining respectively the lower generation masses and 
mixing matrix elements.  

Consider first mixing matrix elements.  Suppose from the scale of the $t$
mass to that of the $b$ mass, the mass matrix has rotated by an angle
$\theta_{tb}$ (Figure \ref{planemix}).   As explained above, one easily 
obtains then the CKM elements as: $V_{tb} = \cos \theta_{tb}$ and 
$|V_{ts}| = |V_{cb}| = \sin \theta_{tb}$.  From the measured values of 
these elements given in the latest databook \cite{databook}, namely:
\begin{equation}
|V_{tb}| = 0.9990 - 0.9993, \ \ |V_{ts}| = 0.035 - 0.043, \ \ 
   |V_{cb}| = 0.037 - 0.043,
\label{VUD}
\end{equation}
one gets thus from each an estimate of the rotation angle, respectively:
\begin{equation}
\theta_{tb} = 0.0374 - 0.0447,\  0.0350 - 0.0430,\  0.0370- 0.0430,
\label{thetatb}
\end{equation}
the values obtained being fully consistent with one another.  (One notes 
that from the same Figure \ref{planemix}, one could deduce in principle 
also $V_{cs} = \cos \theta_{tb}$, but this will be seen, in contrast to 
the 3 other mixing elements already considered, to be a poor approximation 
receiving large nonplanar corrections when all 3 generations are taken 
into account.)

Consider next the second generation masses obtained by the leakage mechanism.
Suppose from the scale of the $t$ mass to that of the $c$ mass, the mass
matrix has rotated by an angle $\theta_{tc}$.   Using (\ref{planelk})
and the
measured values of $m_t$ and $m_c$ given in \cite{databook}, namely:
\begin{equation}
m_t = 174.3 \pm 5.1\ {\rm GeV}, \ \ m_c = 1.15 - 1.35\ {\rm GeV},
\label{mtc}
\end{equation}
one obtains the estimate:
\begin{equation}
\theta_{tc} = 0.0801 - 0.0894.
\label{thetatc}
\end{equation}
Similarly, from the measured values from \cite{databook}:
\begin{equation}
m_b = 4.0 - 4.4\ {\rm GeV}, \ \ m_s = 75 - 170\ {\rm MeV},
\label{mbs}
\end{equation}
one obtains the estimate:
\begin{equation}
\theta_{bs} = 0.1309 - 0.2076,
\label{thetabs}
\end{equation}
the error being so large because of the intrinsic uncertainty in defining
the $s$ quark mass, while from the measured values from \cite{databook}:
\begin{equation}
m_\tau = 1.777\ {\rm GeV}, \ \ m_\mu = 105.66\ {\rm MeV},
\label{mtaumu}
\end{equation}
one obtains the estimate:
\begin{equation}
\theta_{\tau \mu} = 0.2463.
\label{thetaumu}
\end{equation}

\begin{figure}
\centering
\hspace*{-3.5cm}
\input{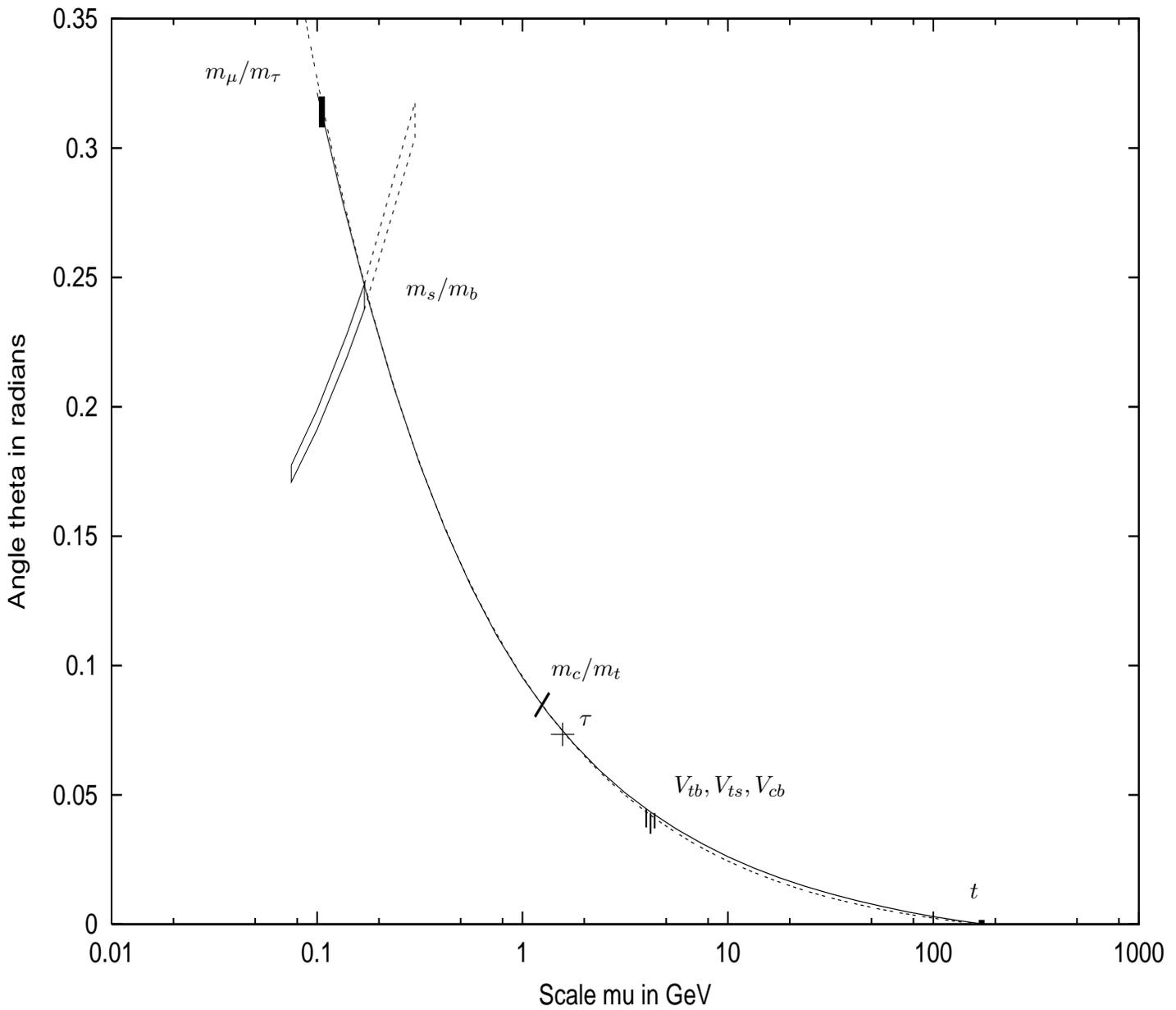}
\caption{The rotation angle changing with scale as extracted from data on
mass ratios and mixing angles and compared with the best fit to the data
(dashed curve) and the earlier calculation by DSM 
(full curve) \cite{phenodsm}, in the planar approximation.}
\label{planero}
\end{figure}

Assume now that the mass matrices of the $U$ and $D$ quarks as well as 
the charged leptons are all aligned at the same scale as proposed above,
and plot the values of the rotation angles obtained before, all starting 
from the direction of the $t$ quark state.  One obtains then the Figure 
\ref{planero} where, rotation angles being simply additive in the planar 
approximation, we have taken $\theta_{ts} = \theta_{tb} + \theta_{bs}$, and 
$\theta_{t \mu} = \theta_{t \tau} + \theta_{\tau \mu}$, with $\theta_{tb}$
taken from (\ref{thetatb}) and $\theta_{t \tau}$ (indicated by a cross
in Figure \ref{planero}) estimated by interpolation between the values 
of $\theta_{tb}$ and $\theta_{tc}$ given above.   One sees first that 
the information gathered so far indeed appear consistent with the data
points lying on a single smooth 
rotation curve as can be explained by a mass matrix rotating smoothly
as the scale changes, which we regard as already a nontrivial support
for the rotation hypothesis independent of any other theoretical
consideration.  In addition, one notes that the data by themselves are
already indicative of the rotation angle approaching an asymptotic
value thus suggesting a rotational fixed point at infinite scale as
predicted by DSM.  Indeed the data points all sit surprisingly well on
the rotation curve obtained a few years earlier \cite{phenodsm} in our
DSM calculation.   Making a best fit to the data by MINUIT 
produces an excellent fit:
\begin{equation}
\theta = \exp(-2.267 - 0.509 \ln \mu) - 0.0075,
\label{bestfit}
\end{equation}
for $\mu$ in GeV, with a $\chi^2$ of 0.21 per degree of freedom, which 
is hardly distinguishable from the DSM curve, as can be seen in the same 
figure.  

The above analysis was performed under the simplifying assumption of
there being only 2 generations of fermion states but one can show that
for the quantities so far considered it is already a good approximation
to the actual 3-generation situation.  When all 3 generations are taken
into account, the mixing matrix can be parametrized as:
\begin{equation}
\left( \begin{array}{ccc} V_{tb} & V_{ts} & V_{td} \\
   V_{cb} & V_{cs} & V_{cd} \\ V_{ub} & V_{us} & V_{ud}
\end{array} \right) =
\left( \begin{array}{ccc} c_1 & -s_1 c_3 & -s_1 s_3 \\
   s_1 c_2 & c_1 c_2 c_3 - s_2 s_3 e^{i \delta} & c_1 c_2 s_3 + s_2 c_3
     e^{i \delta} \\
   s_1 s_2 & c_1 s_2 c_3 + c_2 s_3 e^{i \delta} & c_1 s_2 s_3 - c_2 c_3
     e^{i \delta}
   \end{array} \right),
\label{VKM}
\end{equation}
which is the original Kobayashi-Maskawa parametrization \cite{CKM}, 
only with the ordering of fermion states reversed.  One sees then that if 
we continue to denote as before $V_{tb}$ as $\cos \theta_{tb}$, the elements 
$V_{ts}$ and $V_{cb}$ are no longer just given by $\sin \theta_{tb}$ but 
by respectively $\sin \theta_{tb} \cos \theta_3$ and $\sin \theta_{tb} 
\cos \theta_2$.  The angles $\theta_2$ and $\theta_3$, however, are small
as can be estimated from the empirical values given in \cite{databook} 
for the corner elements of the CKM matrix in comparison to the values of
$V_{ts}$ and $V_{cb}$ quoted above, giving:
\begin{eqnarray}
|V_{td}| = 0.004 - 0.014 & \longrightarrow & |\tan \theta_3| = 0.093 - 
   0.400, \nonumber \\
|V_{ub}| = 0.002 - 0.005 & \longrightarrow & |\tan \theta_2| = 0.047 - 
   0.135,
\label{tan23}
\end{eqnarray}
from which one gets:
\begin{equation}
\cos \theta_2 = 0.999 - 0.991; \ \ \cos \theta_3 = 0.996 - 0.928.
\label{cos23}
\end{equation}
Hence, one concludes that in the 2 generation planar approximation of 
Figures \ref{planemix} and \ref{planero} where one puts $V_{ts} = V_{cb} 
= \sin \theta_{tb}$, one has made an error of at most a few percent which 
is seen hardly to affect the plot or any of the previous remarks we made.
A similar error has been made in Figure \ref{planero} as regards the
lower generation masses obtained from the ``leakage'' mechanism.  The 
estimate (\ref{thetabs}) is for the angle rotated between the scales of 
$m_b$ and $m_s$ but in Figure \ref{planero} we have added this angle
to the rotation angle from scale $m_t$ to scale $m_b$ to get the angle
from scale $m_t$ to scale $m_s$.  Such an addition is valid in the 2
generation approximation but has nonplanar corrections in the actual
3 generation situation.  The error so incurred can be estimated as
follows.  The angle between the plane defined by the $t$ and $c$ vectors 
and the plane defined by $b$ and $s$ is given by the angle between their 
normals, namely the vectors for $u$ and $d$ respectively, which according 
to \cite{databook} takes the value:
\begin{equation}
|V_{ud}| = \cos \theta_{ud} = 0.9742 - 0.9757 
\label{thetaud1}
\end{equation}
giving
\begin{equation}
\theta_{ud} = 0.2209 - 0.2276.
\label{thetaud2}
\end{equation}
The nonplanar error incurred in the angle at scale $m_s$ plotted in Figure 
\ref{planero} is of order $1 - \cos \theta_{ud}$ and is thus of order again 
a few percent, which is negligible given the large error already inherent 
in the definition of the $s$ quark mass.  A similar error is presumably 
present in the angle plotted in Figure \ref{planero} at scale $m_\mu$.
More details of this analysis can be found in
a preliminary report \cite{empirdsm}.

However, though good for illustration purposes because of its simplicity,
the above analysis in the planar approximation is incomplete in that it
is restricted only to scales above the $\mu$ mass where nonplanar effects
are negligible, and hence cannot account for all the available empirical 
information which includes the data from neutrino oscillations at very low
scales where nonplanar effects can no longer be neglected.  Furthermore, 
to test the rotation hypothesis exhaustively one has also to ensure that 
no hidden violation of the hypothesis exists in the off-planar direction.  
In any case, the planar approximation, though good for
illustrative purposes, actually need not be made 
since, as detailed 
in (\ref{mfact}) above, the whole content of the rotating mass matrix 
even when all 3 generations are taken into account is encapsulated just 
in the single 3-dimensional vector ${\bf r}(\mu)$ depending on scale $\mu$.  
The only technical problem posed by a full analysis is thus how to 
extract this vector at various scales from the existing data on fermion 
mass ratios and mixing parameters.  Once so extracted, the vector can 
be confronted with the rotation hypothesis and should be consistent with 
tracing out a continuous curve in 3-space if the hypothesis is correct.  
For the extraction of this vector, we now proceed as follows.

\section{Extracting ${\bf r}(\mu)$ from quark data}

The $U$-quarks $t, c, u$ are independent quantum states so that their 
state vectors should form an orthonormal triad in generation space, which
we can choose without loss of generality as:
\begin{equation}
{\bf v}_t = (1,0,0); \ \ {\bf v}_c = (0,1,0); \ \ {\bf v}_u = (0,0,1). 
\label{Utriad}
\end{equation}
The $D$-quark state vectors also form an orthonormal triad the orientation
of which relative to the $U$-triad is given by the CKM matrix elements:
\begin{equation}
{\bf v}_b = (V_{tb},V_{cb},V_{ub}); \ \ {\bf v}_s = (V_{ts},V_{cs},V_{us}); 
   \ \ {\bf v}_d = (V_{td},V_{cd},V_{ud}). 
\label{Dtriado}
\end{equation}
Hence, if the complex elements of the CKM matrix are accurately known, the
$D$-triad would also be determined.  At present, however, only the absolute 
values of the CKM matrix elements are experimentally known to reasonable
accuracy, leading thus to some ambiguities in the determination of the 
$D$-triad.  In particular, one is forced to ignore for the moment in the
CKM matrix the CP-violating phase which is experimentally very poorly 
determined and treat the $D$-triad also as real vectors.  Inserting then the 
experimental limits on the CKM matrix elements as read from \cite{databook} 
gives rather tight constraints on the directions of the $D$-triad
with errors so small as to be mostly negligible for our analysis:
\begin{eqnarray}
{\bf v}_b & = & ((0.9990-0.9993), (0.037-0.043), -(0.002-0.005)); \nonumber \\
{\bf v}_s & = & (-(0.035-0.043), (0.9734-0.9749), (0.219-0.226)); \nonumber \\
{\bf v}_d & = & ((0.004-0.014), -(0.219-0.225), (0.9742-0.9757)).
\label{Dtriad}
\end{eqnarray}
The signs of the 3 components for $b$ can be chosen arbitrarily by
choosing the physically irrelevant phases of the various quark state 
vectors and a particular choice has been made in (\ref{Dtriad}) for 
convenience.  The signs for the other 2 states are then determined by 
orthogonality.  Actually, given the present errors on the CKM matrix 
elements, there is an alternative solution to that shown for the state 
vectors of $s$ and $d$, which however we can ignore for reasons  
to be explained below. 

We notice that by the considerations in the preceding section, the state 
vector of $b$, this being the heaviest state in the $D$ sector, is just 
the rotating vector ${\bf r}(\mu)$ taken at the scale $\mu = m_b$, thus:
\begin{equation}
{\bf r}(m_b) = {\bf v}_b.
\label{rmb}
\end{equation}
Together with:
\begin{equation}
{\bf r}(m_t) = {\bf v}_t = (1,0,0),
\label{rmt}
\end{equation} 
we have then two points on the trajectory for ${\bf r}(\mu)$ we wish to 
trace.  One convenient way to present this, we find, is to write this
rotating vector as ${\bf r}(\mu) = (\xi(\mu),\eta(\mu),\zeta(\mu))$ with
$\xi(\mu)^2 = 1-\eta(\mu)^2 - \zeta(\mu)^2$ and plot the 
constraints on
$\eta(\mu)$ and $\zeta(\mu)$ on the $\eta \zeta$-plane.  The result from 
(\ref{rmt}) and (\ref{rmb}) are then entered as the first 2 points from 
the left in the 3-D plot of Figure \ref{3Dplot}.  This plot, which shows
$\eta(\mu)$ and $\zeta(\mu)$ as functions of the energy scale $\mu$ can in 
principle incorporate all the information that we shall extract from data.  
However, it being often hard to read the information it contains, we shall 
supplement it by its 3 projections onto the 3 co-ordinate planes, namely 
onto the $\eta\zeta$-plane in Figure \ref{etazeta}, the $\mu\eta$-plane 
in Figure \ref{mueta}, and the $\mu\zeta$-plane in Figure \ref{muzeta}, 
which projections, as we shall see, will be useful later also for 
interpolation and extrapolation purposes. 

\begin{figure*}
\vspace*{-3cm}
\centering
\includegraphics[scale=0.9]{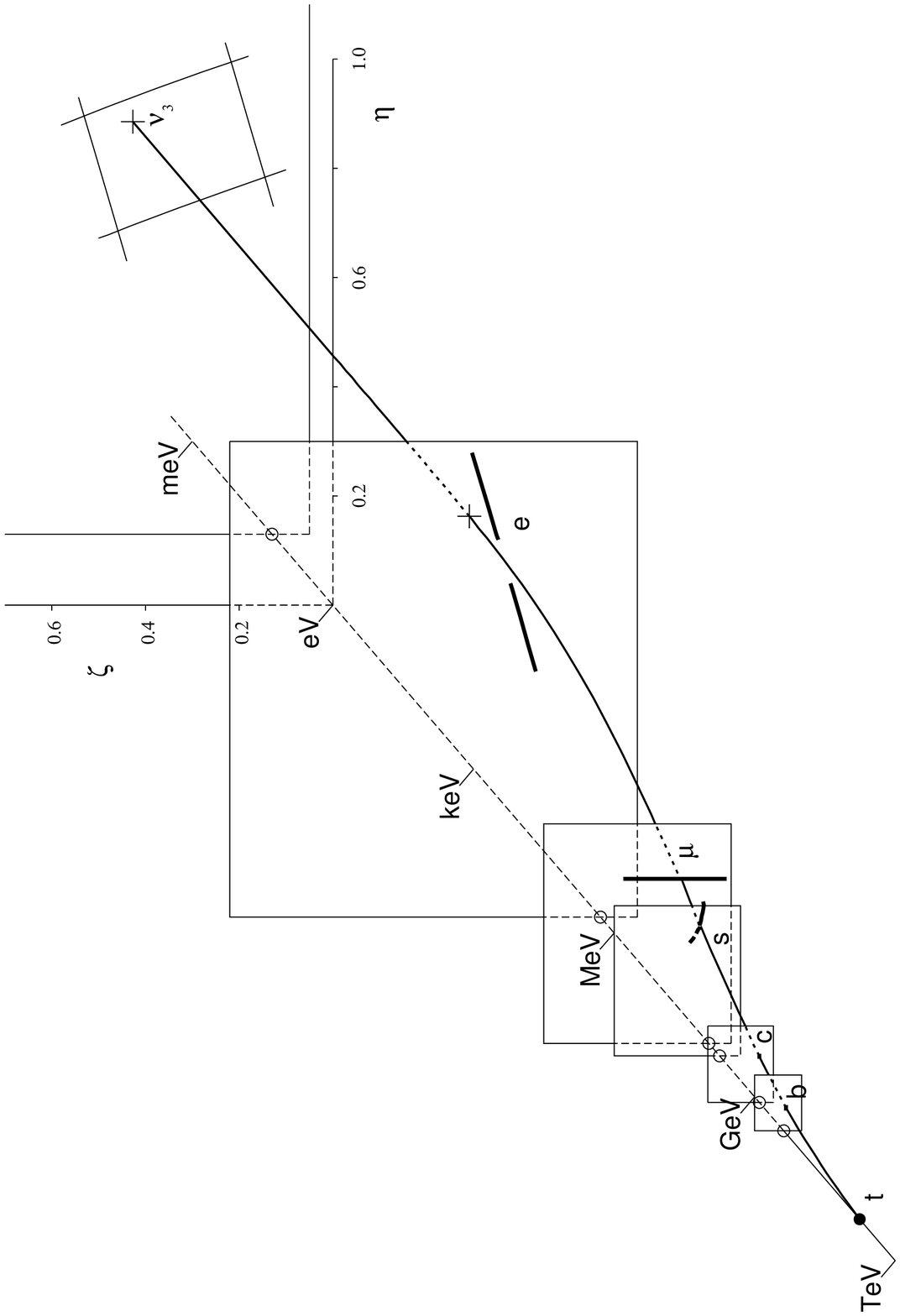}
\end{figure*}

\begin{figure}
\vspace{8cm}
\caption{A plot of the rotating vector ${\bf r}(\mu)$ as extracted from 
existing data on fermion mass ratios and mixing parameters, where its 
second and third
components, i.e.\ $\eta(\mu)$ and $\zeta(\mu)$, are plotted as 
functions of $\ln \mu$, $\mu$ being the energy scale.  The experimentally
allowed values at any one scale are represented as an allowed region on a
plaquette, with the scale corresponding to a plaquette being given by the
intersection, denoted by a small circle, of its left-most boundary with 
the $\mu$-axis.  For example, the first small plaquette on the left of the 
figure corresponds to the scale $\mu = m_b$, on which plaquette the allowed 
region for ${\bf r}(\mu) = {\bf v}_b$ is very small because of the small
experimental error on the CKM matrix elements $V_{tb}, V_{cb}$ and $V_{ub}$. 
The last plaquette on the right, on the other hand, corresponds to the
scale $\mu = m_{\nu_3}$, on which plaquette the allowed region for 
${\bf r}(\mu)$ is a rough rectangular area bounded by the data on $\nu$ 
oscillations from atmospheric neutrinos and from the Chooz experiment.  The 
curve represents the result of a DSM one-loop calculation from an earlier 
paper \cite{phenodsm} which is seen to pass through the allowed region on 
every plaquette except that for the electron $e$.  For further explanation 
of details, please see text.} 
\label{3Dplot}
\end{figure}

\begin{figure}
\vspace*{-3cm}
\centering
\includegraphics[scale=0.8]{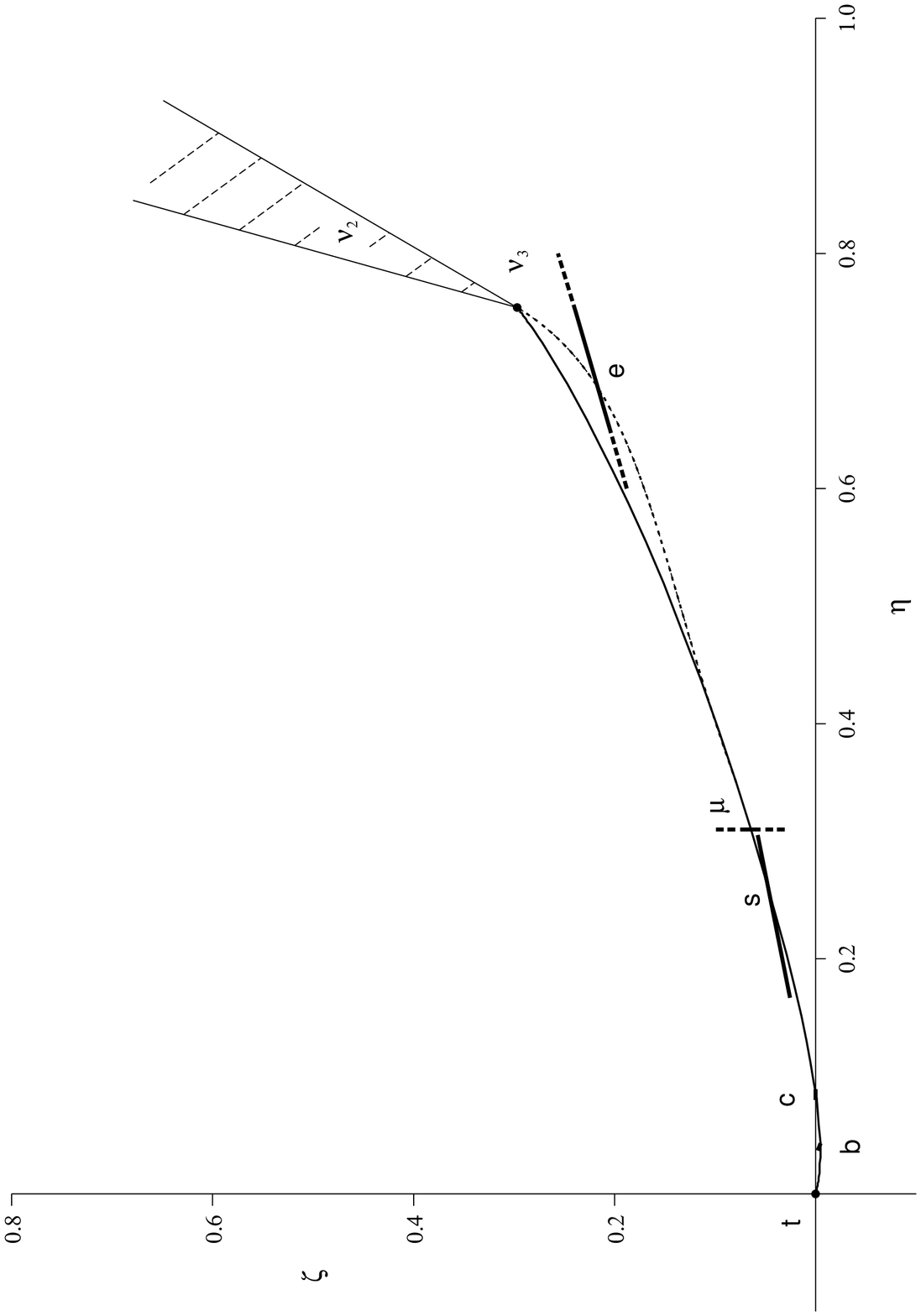}
\caption{Projection of Figure \ref{3Dplot} onto the $\eta\zeta$-plane.
The full curve represents the DSM one-loop calculation of \cite{phenodsm} 
and the dashed curve its suggested deformation at low scales to fit the 
data on $m_e$ and $U_{e2}$.}
\label{etazeta}
\end{figure}

\begin{figure}
\vspace*{-3cm}
\centering
\includegraphics[scale=0.8]{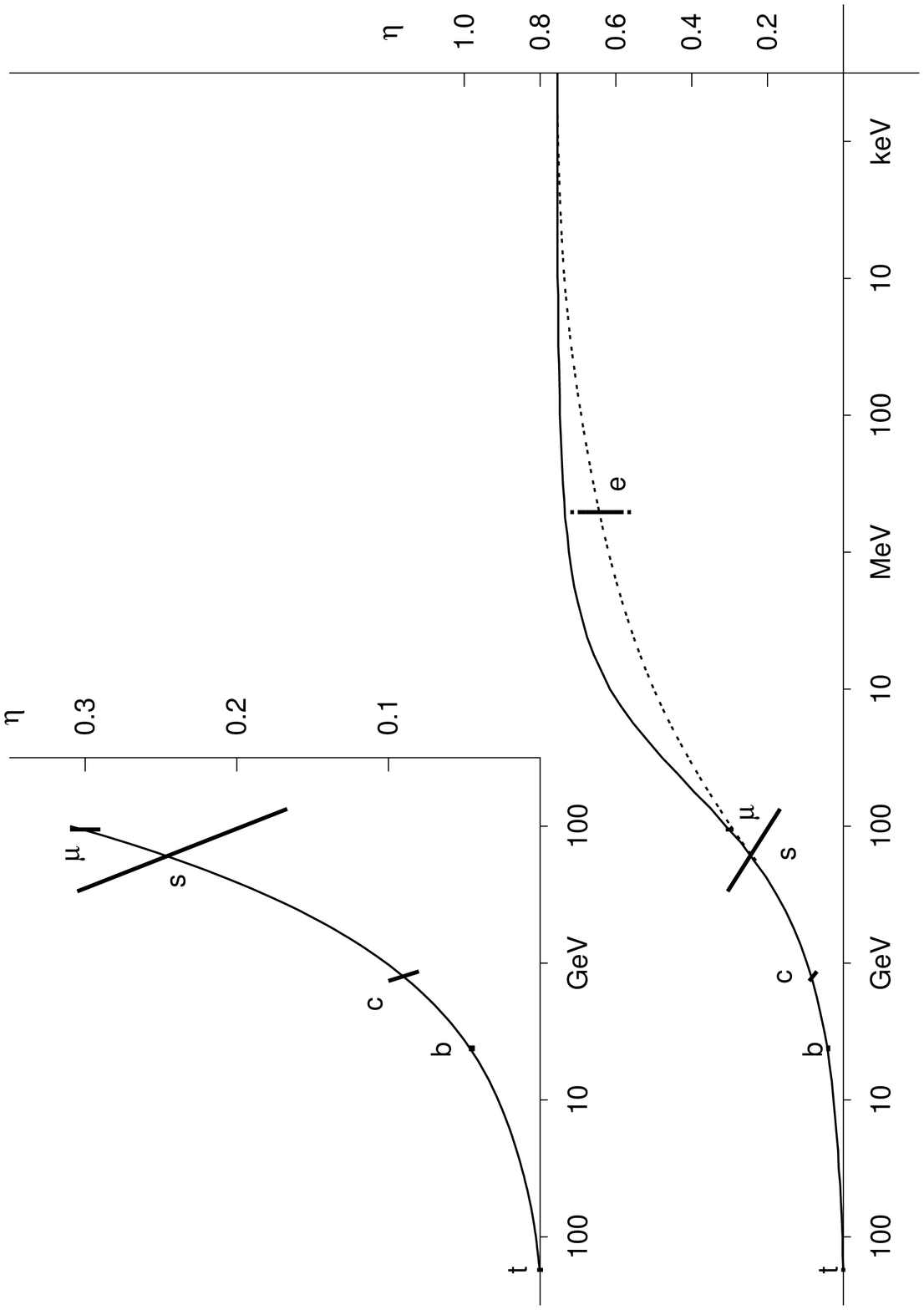}
\caption{Projection of Figure \ref{3Dplot} onto the $\mu\eta$-plane.
The full curve represents the DSM one-loop calculation of \cite{phenodsm} 
and the dashed curve its suggested deformation at low scales to fit the 
data on $m_e$ and $U_{e2}$.}
\label{mueta}
\end{figure}

\begin{figure}
\vspace*{-3cm}
\centering
\includegraphics[scale=0.8]{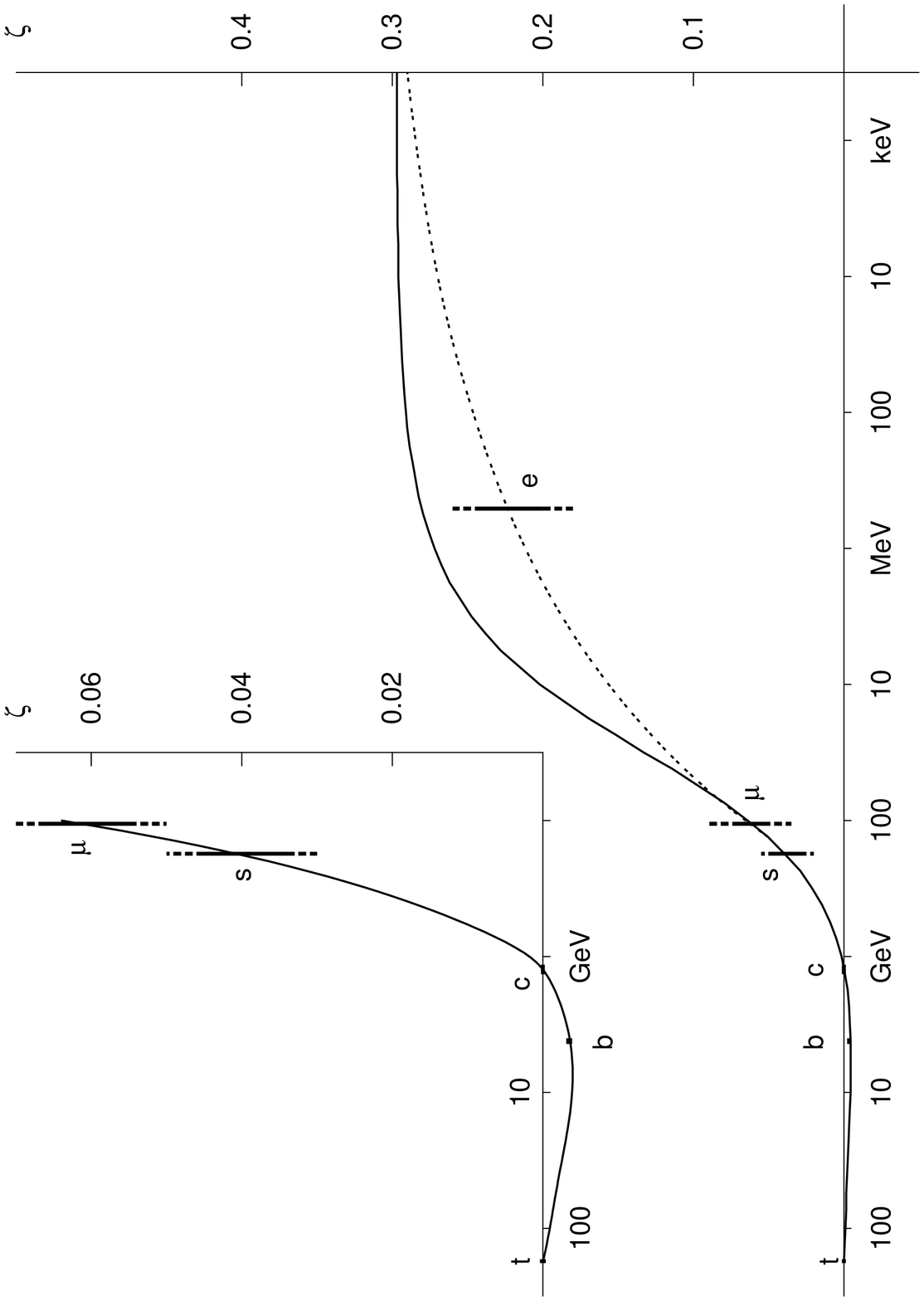}
\caption{Projection of Figure \ref{3Dplot} onto the $\mu\zeta$-plane.
The full curve represents the DSM one-loop calculation of \cite{phenodsm} 
and the dashed curve its suggested deformation at low scales to fit the 
data on $m_e$ and $U_{e2}$.}
\label{muzeta}
\end{figure}

Next, we turn to consider ${\bf r}(\mu)$ at the scale $\mu = m_c$.  Since
the mass of $m_c$, by our original hypothesis, comes about only through 
the ``leakage'' from $m_t$ as detailed above, it follows that the vector 
${\bf r}(m_c)$ will have to be a linear combination: $\cos \theta_{tc} 
{\bf v}_t + \sin \theta_{tc} {\bf v}_c$, of the state vectors ${\bf v}_t$ 
and ${\bf v}_c$ of respectively $t$ and $c$ with $\sin \theta_{tc}$ given 
by (\ref{planelk}), or explicitly: 
\begin{equation}
{\bf r}(m_c) = \sqrt{1 - m_c/m_t}\;{\bf v}_t + \sqrt{m_c/m_t}\;{\bf v}_c.
\label{ratmc}
\end{equation}
Inputting the quoted experimental limits on $m_t$  and
$m_c$ gives us the 
allowed region for the third data point on the trajectory of ${\bf r}(\mu)$, 
as is shown on Figure \ref{3Dplot}.  There is in fact of course another 
solution for ${\bf r}(m_c)$ in (\ref{ratmc}) with $\sin \theta_{tc}
= - \sqrt{m_c/m_t}$, meaning that there is an additional disjoint
branch of the allowed region in another quadrant of the $\eta
\zeta$-plane to that displayed in the figure.  The existence of this
additional branch, however, does not affect the question of interest
to us here, namely, whether the allowed region is consistent with data
lying on a smooth rotation curve, so long as the first branch already
does, and can therefore be ignored.  Such additional branches of the
allowed regions will in fact always occur for all other pieces of
information on ${\bf r}(\mu)$ that we shall extract from data, since
experiment so far gives only the absolute values of the relevant
quantities but, for the same reason as above, these additional
disjoint branches for our present purpose can almost always be ignored.

The next point in line is ${\bf r}(\mu)$ at $\mu = m_s$, which will of 
necessity be poorly determined because the $s$ mass is very poorly known.
Nevertheless, whatever is taken for the mass of $s$ so long as it is given
by the ``leakage mechanism'' from $b$, ${\bf r}(m_s)$ will have to be a 
linear combination of the state vector ${\bf v}_b$ of $b$ and the state 
vector ${\bf v}_s$ of $s$, so that by reasoning exactly as above for $c$, 
we have:
\begin{equation}
{\bf r}(m_s) = \sqrt{1 - m_s/m_b}\;{\bf v}_b + \sqrt{m_s/m_b}\;{\bf v}_s.
\label{ratms}
\end{equation}
The range of values for $m_s$ is given in \cite{databook} as 75---170 MeV, 
and in \cite{databook96} as 100---300 MeV, depending on the scale at which
the limits were determined respectively.  The allowed region shown for 
${\bf r}(m_s)$ in Figure \ref{3Dplot} corresponds to the union of the 
above two ranges for $m_s$ values, which allowed region, one will notice, 
does not lie entirely on the plaquette corresponding to  an $m_s$ 
value at 176 MeV but protrudes to either side of it.  This gives then 
a 4th point on 
the trajectory for ${\bf r}(\mu)$.  As noted before, given the present 
errors on the CKM mattrix elements, there is actually an alternative 
solution for ${\bf v}_s$ with a different sign for the third component   
to that given in (\ref{Dtriad}), meaning an additional branch to the
allowed region for ${\bf r}(m_s)$, but this can be ignored for the
same reason as that given for ${\bf r}(m_c)$ above.

The above represents more or less the most that can be extracted from the
data on quark masses and mixing (barring CP violation) about the rotating
vector ${\bf r}(\mu)$ apart from some as yet rather uncertain information
from the masses of the light quarks $u$ and $d$.  The method we used for
extracting the fermion masses from the ``leakage mechanism'' as contained
in e.g.\ (\ref{planelk}) was meant only for freely propagating particles
and should not in principle be applied to quarks which are confined, except 
approximately to the heavier quarks which are generally regarded as 
quasi-free.  For the light quarks $u$ and $d$ which are tightly confined, 
it is clearly not applicable.  The question then arises in what way these
light quark masses are to be defined.  Experimentally, these masses are
determined at some convenient but somewhat arbitrary scale
such as  1 or 2 GeV,
and it is not clear what these values should correspond to in the ``leakage
mechanism''.  One possibility is to consider these mass values as the 
``leakage'' from the rotating vector ${\bf r}(\mu)$ taken at the chosen 
scale 1 or 2 GeV into the vectors ${\bf v}_u$ and ${\bf v}_d$ thus:
\begin{equation}
|{\bf v}_u.{\bf r}(\mu)|^2 \stackrel{?}{=} 
m_u/m_t; \ \ |{\bf v}_d.{\bf r}(\mu)|^2 \stackrel{?}{=} m_d/m_b.
\label{mumdleak}
\end{equation}
In that case one obtains values for both $m_u$ and $m_d$ of the right order
of magnitude in the MeV region, but it is not certain whether this is of much 
significance.  As matters stand, therefore, we can only leave open the 
question of the light quark masses.

\section{Extracting ${\bf r}(\mu)$ from lepton data}
 
To proceed further, one turns now to the leptons for which the preceding 
analysis for quarks can in principle be independently repeated, for as far 
as the premises of the rotating mass matrix as set up at the beginning of 
this paper is concerned, there is strictly nothing which needs connect
the mass matrices of quarks and leptons.  However, the DSM scheme suggests
that the rotating matrices for quarks and leptons both lie on the same
trajectory, the last being specified just by the vev's of the (dual colour)
Higgs fields which are independent of the fermion type \cite{ckm,phenodsm}, 
and this suggestion has been borne out by the 2-generation analysis
reported above.  
It makes practical sense 
therefore to adopt the same position here, especially since it raises the 
stakes and makes the present analysis an even more stringent test for the 
rotation hypothesis.  This means in particular that the state vector, say
${\bf v}_\tau$, of the $\tau$ lepton, this being the heaviest eigenstate 
of the lepton mass matrix, could be identified again with the vector 
${\bf r}(\mu)$ taken at the scale $\mu = m_\tau$, the location of which 
can readily be determined by interpolating 
${\bf r}(\mu)$ between $m_b$ and $m_c$, as in Figure~\ref{planero}.

Having fixed ${\bf v}_\tau$, one can constrain the vector ${\bf r}(\mu)$ 
at $\mu = m_\mu$ by the condition:
\begin{equation}
|{\bf r}(m_\mu).{\bf v}_\mu|^2 = m_\mu/m_\tau,
\label{rmmu}
\end{equation}
with:
\begin{equation}
{\bf v}_\mu.{\bf v}_\tau = 0,
\label{vmuovtau}
\end{equation}
or in other words:
\begin{equation}
1 - |{\bf r}(m_\mu).{\bf v}_\tau|^2 = m_\mu/m_\tau.
\label{mmu}
\end{equation}
This gives 4 solutions for $\eta(m_\mu),\zeta(m_\mu)$, corresponding to 
respectively the two signs of ${\bf r}(m_\mu).{\bf v}_\tau$ and the two 
signs of $\xi(m_\mu) = \pm\sqrt{1-\eta(m_\mu)^2 - \zeta(m_\mu)^2}$.  These 4
solution are, however, widely separated, so that only the solution
with ${\bf r}(m_\mu).{\bf v}_\tau$ and $\xi(m_\mu)$ both positive is
shown in Figure \ref{3Dplot} which, 
by inputting the empirical values
of $m_\tau = 1777 \ {\rm MeV}$ and $m_\mu = 105 \ {\rm MeV}$ taken from 
\cite{databook}, gives as the allowed region for ${\bf r}(m_\mu)$ a narrow 
band on the $\mu$ plaquette approximately parallel to the $\zeta$-axis, 
the width of the band representing the error on ${\bf v}_\tau$ obtained 
from the above interpolation.  For the rotation hypothesis to be valid, 
the trajectory for ${\bf r}(\mu)$ is required to pass through this band 
at $\mu = m_\mu$ .

The above information on the vector ${\bf r}(m_\mu)$ determines also 
to a fair approximation the state vector ${\bf v}_\mu$, the latter being 
constrained by the ``leakage'' mechanism to lie on the plane containing 
${\bf v}_\tau$ and ${\bf r}(m_\mu)$ and to be orthogonal to ${\bf v}_\tau$.   
That this is so can be seen as follows.  As noted already, the vector 
${\bf v}_\tau$ being near the vector ${\bf r} (m_c)$ and therefore
lying very nearly on the $\xi\,\eta$-plane, the 
allowed band for ${\bf r}(m_\mu)$ defined by (\ref{mmu}) is very nearly 
parallel to the $\zeta$-axis so that the second component of ${\bf r}(m_\mu)$, 
namely $\eta(m_\mu)$, is very well determined, as is depicted in Figure 
\ref{mueta}.  By interpolating with a curve drawn through the four quite
accurate points for respectively $t, b, c$ and $\mu$, one can then get a 
fair estimate for the value of $\eta(m_s)$.  Hence from Figure \ref{etazeta},
one can read off the corresponding value for the third component $\zeta(m_s)$ 
which on insertion into Figure \ref{muzeta} then allows for
an extrapolation
to the $\mu$ mass scale to give an estimate for the value of $\zeta(m_\mu)$, 
which though rough, being in any case small, is sufficient for our purpose.
Having then obtained the vector ${\bf r}(m_\mu)$, the state vector of $\mu$,
namely ${\bf v}_\mu$, is also determined by the conditions stated at the
beginning of the paragraph.  At the same time, of course, the state vector
${\bf v}_e$ of $e$ is also determined by orthogonality to both ${\bf v}_\mu$
and ${\bf v}_\tau$.  The actual numerical values we so obtained for the
charged leptonic triad which we shall use later for our analysis are as 
follows:
\begin{eqnarray}
{\bf v}_\tau & = & (0.9975, 0.0700, -0.0015), \nonumber \\
{\bf v}_\mu  & = & (-0.0654, 0.9516, 0.3003), \nonumber \\
{\bf v}_e    & = & (0.0224, -0.2995, 0.9538).
\label{Ltriad}
\end{eqnarray}
This determination of the charged lepton triad on which the analysis in 
the remainder of this section depends is about the best that one can 
do for the moment but is obviously not as accurate as one could wish.  
Nevertheless, as we shall see, it still serves its purpose in allowing 
us to extract some interesting information on ${\bf r}(\mu)$ for the 
low $\mu$ region.

First, according to the ``leakage mechanism'', the mass of the electron is
given by:
\begin{equation}
|{\bf r}(m_e).{\bf v}_e|^2 = m_e/m_\tau,
\label{rme}
\end{equation}
which, as for (\ref{rmmu}) and for the same reasons, gives 4 solutions, 
2 of which corresponding to negative values for $\xi(m_e)$ lie outside
Figure~\ref{3Dplot},
leaving two which are 
represented  by respectively the line drawn on 
the $e$-plaquette and another line (not shown) nearly parallel to the
first but 0.035 units lower (and hence almost coinciding with the one 
shown).  
Again, for the rotation hypothesis to be 
valid, the trajectory for ${\bf r}(\mu)$ has to pass through one of 
these 2 line at $\mu = m_e$.  In obtaining these lines, we have of
course input the well known value of 0.51 MeV for the mass of the 
electron.
 
Secondly, the MNS \cite{MNS} lepton mixing matrix elements $U_{\mu3}$ 
and $U_{e3}$, as studied in oscillation experiments on respectively 
atmospheric neutrinos \cite{superK,Soudan} and reactor neutrinos such 
as \cite{Chooz}, are given by the inner products:
\begin{equation}
U_{\mu3} = {\bf v}_\mu.{\bf v}_3,  
\label{Umu3}
\end{equation}
and:
\begin{equation}
U_{e3} = {\bf v}_e.{\bf v}_3,  
\label{Ue3}
\end{equation}
with ${\bf v}_3 = {\bf r}(m_{\nu_3})$ being the state vector of the 
heaviest neutrino $\nu_3$.  With the $\mu$ state vector as determined
above in (\ref{Ltriad}), one obtains by inputting the experimental range 
for $|U_{\mu3}|^2$ of about 1/3 to 2/3 \cite{superK,Soudan} again four 
solutions for the allowed region, three of which lie outside
Figure~\ref{3Dplot},
leaving one (corresponding to $U_{\mu3}$ and
$\xi(m_{\nu_3})$ both positive) 
which is represented in the figure by 
the area bounded by the two near vertical lines on the $\nu_3$-plaquette.  
Similarly, with the $e$ state vector as determined in (\ref{Ltriad}), 
one obtains by inputting the experimental bound $|U_{e3}|^2 < 0.027$ 
\cite{Chooz}, four solutions for the allowed region, but this time only 
two (corresponding to $\xi(m_{\nu_3})$ negative)
can be ignored being outside Figure~\ref{3Dplot}, the other two 
being adjacent merge into one as represented in the figure by 
the area bounded by the two near horizontal lines on the $\nu_3$-plaquette. 
The consequent allowed region for the vector ${\bf r}(m_{\nu_3})$ is thus
represented by the roughly rectangular area 
shown, where we have put $m_{\nu_3}^2
\sim 3 \times 10^{-3} {\rm eV}^2$ as preferred by \cite{superK,Soudan}
and \cite{K2K}.

Finally, the mixing element $U_{e2}$ as inferred from solar neutrino
experiments is given as:
\begin{equation}
U_{e2} = {\bf v}_e.{\bf v}_2,  
\label{Ue2}
\end{equation} 
where ${\bf v}_2$ is the state vector for the second heaviest neutrino
$\nu_2$ which is by definition orthogonal to ${\bf v}_3$.  Following thus
the same procedure as in the preceding paragraph, one can determine the 
allowed region for the vector ${\bf v}_2$ by inputting the bounds on
${\bf v}_3$ as obtained above and the experimental bounds on $U_{e2}$ 
\cite{superK,Sno,Fogli}.  However, to extract ${\bf r}(m_{\nu_2})$ from
this, one would need $m_{\nu_2}$ which is experimentally still largely
unknown so that the above information on ${\bf v}_2$ cannot 
readily be presented
in Figure \ref{3Dplot}.  But, as we shall see, there is another way of 
displaying this information.

\section{Discussion}

The allowed regions for the vector ${\bf r}(\mu)$ for various scales
$\mu$ displayed in Figure~\ref{3Dplot} and its projections 
Figures \ref{etazeta}--\ref{muzeta},
represent all the information on ${\bf r}(\mu)$
that can be extracted at present from 
fermion mass and mixing parameters, apart from the $u,d$ masses and the 
solar neutrino angle $U_{e2}$ already noted.  
This information was extracted on the assumption that both fermion 
mixing and lower 
generation masses arose solely as consequences of the rotation of the 
mass matrix, under which circumstances the rotation is encapsulated 
entirely in the rotating vector ${\bf r}(\mu)$, as was explained in 
equation (\ref{mfact}) or (\ref{mfactex}).  The definition of masses, 
state vectors, and mixing matrices,
which for a rotating mass matrix is delicate, followed the prescription
given in the introduction, which seems to us the natural one and is to 
our knowledge the only one available in the literature.  Apart from 
these, no other theoretical input or assumption has been introduced.
Hence if the rotation hypothesis set out above is correct, then the
allowed regions 
should line up along some smooth 3-D curve from the heaviest $t$ to 
the lowest $\nu_3$.  
Indeed this is seen in the above-quoted figures to be 
the case.  In the high energy region, say down to the $\mu$ 
lepton mass 
scale, where the allowed regions are mostly small, the alignment is seen 
to be quite accurate, not only in the projection of Figure \ref{mueta} 
on to the $\mu\eta$-plane as already noted in Section 2, but 
also in the two other directions as seen in Figures \ref{etazeta} and 
\ref{muzeta}.  Below the 
$\mu$ lepton mass scale, the allowed regions are larger 
and the constraints not too stringent but they are seen still to be 
thoroughly consistent with alignment on a smooth trajectory spanning 
some 13 orders of magnitude in energy.  This looks to us nontrivial
and lends direct empirical support to the rotation hypothesis 
which is entirely model-independent and free from
extraneous theoretical bias.  This is the main conclusion we set out
to establish.

Next, we turn to consider the shape of the trajectory traced out by the 
data in Figure~\ref{3Dplot}, to understand which more theoretical
input will of course be needed.  We shall do so with reference 
to our DSM scheme where 
the whole rotation idea was first proposed and in which a perturbative 
method for calculating the rotating trajectory was suggested and carried 
out already to 1-loop order \cite{ckm,phenodsm}.  First, we notice
from Figures \ref{mueta} and \ref{muzeta} that ${\bf r}(\mu)$ seems 
to be approaching asymptotic 
limits for both $\mu \rightarrow \infty$ and $\mu \rightarrow 0$, and thus 
indicative of rotation fixed points at these scale values.  Now, these
fixed points are predicted by the DSM scheme, and they occur there 
by virtue only of the inbuilt mechanism in the model for driving the 
rotation and are thus independent of the adjustable parameters of the 
model.  The detailed shape of the trajectory, however, does depend on
the parameters of the model, of which there are 3, with 2 giving the
initial direction of the trajectory and 1 other governing the rotation
speed.  The calculation done already a few years ago
\cite{phenodsm} with the 3 parameters in the 
model fitted to the mass ratios $m_c/m_t$, $m_\mu/m_\tau$ and to the Cabibbo 
angle is reproduced in Figures \ref{3Dplot}--\ref{muzeta}.  Although this
result has never been explicitly presented before, it can be inferred from
e.g.\ Figure 3 of \cite{phenodsm} and transformed to the present frame  
(\ref{Utriad}) from the frame used there through the vectors:
\begin{eqnarray}
{\bf v}_t & = & (0.9999,0.0117,0.0008); \nonumber \\
{\bf v}_c & = & (-0.0110,0.9148,0.4038); \nonumber \\
{\bf v}_u & = & (0.0040,-0.4038,0.9149) 
\label{Utriadp}
\end{eqnarray}
obtained from the previous calculation \cite{phenodsm}.  It is seen
to agree very well with the newly extracted information down to
the $\mu$ mass scale.  In particular, the rotational fixed point
predicted by the DSM at $\mu = \infty$ is seen to be fully consistent
with the data.
Below the $\mu$ mass the DSM curve calculated
to 1-loop order begins to deviate from the regions allowed by experiment.
For example, on the $e$-plaquette in Figure \ref{3Dplot}, the DSM curve
if exact should hit the allowed line at $\mu = m_e$ but, as indicated
by the little cross, it hits the plaquette instead at some distance from
the allowed line.  This deviation represents the difference in the mass
of the electron as predicted by the old calculation \cite{phenodsm} from
its true value, i.e.\ 6 MeV instead of 0.51 MeV.  Such a deviation is of 
course expected, since at lower scales, the vector ${\bf r}(\mu)$ moves 
further and further from the high energy fixed point predicted by the 
scheme so that the 1-loop calculation for the trajectory will become 
less and less reliable.  However, the 1-loop approximate trajectory from 
\cite{phenodsm} still hits the $\nu_3$ plaquette inside the 
allowed region, in other words giving correct predictions for the MNS 
mixing elements $U_{\mu3}$ and $U_{e3}$.  This is because these
elements depend only on the vector ${\bf v}_3 = {\bf r} (m_{\nu_3})$
which, as indicated in Figures \ref{mueta} and \ref{muzeta}, is
already near the asymptotic value.  Hence, the fact the the
calculation agrees with data for $U_{\mu3}$ and $U_{e3}$ suggests that
the rotational fixed point at $\mu=0$ is correctly predicted, although
the rotational curve itself near this fixed point is not, by the
1-loop approximation.

In contrast, the state vector ${\bf v}_2$ of the second heaviest neutrino 
$\nu_2$ represents the tangent vector to the trajectory near the low
energy fixed point and cannot therefore be expected to be accurately
predicted by the 1-loop calculation of \cite{phenodsm}.  Indeed, the
value predicted by \cite{phenodsm} for the mixing element $U_{e2}$ which
depends on ${\bf v}_2$ fell outside the limits set by the solar neutrino 
experiments.  In our present analysis, the information on ${\bf v}_2$
extracted from the experimental limits on $U_{e2}$ can be presented as
a wedge-shaped region in Figure \ref{etazeta} in which the tangent to
the trajectory at the low energy fixed point is supposed to lie, which 
region is estimated with a bound $|U_{e2}|^2 \sim 0.33 \pm 0.1$ favoured by
present experiments \cite{superK,Sno}.  As can be seen in the figure, 
the trajectory predicted by the DSM 1-loop calculation does not satisfy 
this criterion.  Again, as in previous cases, there are in fact four 
solutions to this allowed region, among which we have chosen 
to display the one which is nearest to accommodating the DSM 1-loop 
trajectory.  However, this is not surprising since it is already expected 
that the 1-loop trajectory will be unreliable below the $\mu$ mass scale.  
In that case, it may be interesting turning the argument around to use 
the information at low scale, scanty though it is at present, to constrain 
the exact trajectory if such really exists.  One sees then that just by 
deforming somewhat the 1-loop curve, one would be able to remove both the 
previously noted discrepancies in the $e$ mass and in the mixing element 
$U_{e2}$, as indicated in Figures \ref{etazeta}, \ref{mueta} and 
{\ref{muzeta}.  

In summary, we conclude that the existing data on fermion mass and mixing 
when appropriately interpreted do support the hypothesis of a mass matrix
rotating with changing scales, and that the rotation trajectory indicated 
bears a close resemblance to that predicted earlier by the DSM scheme.

We thank Carmen Garcia Garcia for kindly helping us with the best
fit to the data presented in Figure~\ref{planero}, and Bill Scott for
advising us on the neutrino oscillation data.


\begin{thebibliography}{99}

\bibitem{CKM}  N.\ Cabibbo, Phys.\ Rev.\ Lett.\ {\bf 10}, 531 (1963);
   M.\ Kobayashi and T.\ Maskawa, Prog.\ Teor.\ Phys.\ 49, 652 (1973).

\bibitem{dualgen}  See e.g. Chan Hong-Mo hep-th/0007016, Int. J. Mod. 
   Phys, A16, 163, (2001), and/or Chan Hong-Mo and Tsou Sheung Tsun, 
   hep-ph/0008312 (2000), Proc. 8th Asia-Pacific Phys. Conf. APPC 2000,
   ed. Yeong-Der Yao et al. (World Scientific, Singapore 2001) p. 447.

\bibitem{MNS}  Z. Maki, M. Nakagawa and S. Sakata, Progr. Theor. Phys. 
   28, 870 (1962).

\bibitem{Jarlskog} C. Jarlskog, in {\it CP Violation}, ed. C. Jarlskog,
   World Scientific, Singapore, 1989.

\bibitem{Weinberg} Steven Weinberg, Phys. Rev. D7, 2887, (1973).

\bibitem{Ramon} H. Arason, D.J. Casta\~no, B. Kesthelyi, S. Mikaelian,
   E.J. Piard, P. Ramond, and B.D. Wright, Phys. Rev. D46, 3945, (1992).

\bibitem{physcons} Chan Hong-Mo and Tsou Sheung Tsun, Phys. Rev. D57,
   2507, (1998).

\bibitem{databook}  Review of Particle Physics, D.E. Groom et al., Eur.
   Phys. Journ. C15, 1, (2000).

\bibitem{phenodsm}  Jos\'e Bordes, Chan Hong-Mo and Tsou Sheung Tsun,
   hep-ph/9901440, Eur. Phys. J. C 10, 63, (1999).

\bibitem{empirdsm}  Jos\'e Bordes, Chan Hong-Mo and Tsou Sheung Tsun,
   hep-ph/0104036 (2001).

\bibitem{databook96}  R.\ M.\ Barnett et.\ al, Phys.\ Rev.\ D54, 1 (1996).

\bibitem{ckm}  Jos\'e Bordes, Chan Hong-Mo, Jacqueline Faridani, Jakov 
   Pfaudler, and Tsou Sheung Tsun,  Phys. Rev. D58, 013004, (1998), 
   hep-ph/9712276. 

\bibitem{superK}  Superkamiokande data, see e.g. talk by T. Toshito at
   ICHEP'00, Osaka (2000).

\bibitem{Soudan}  Soudan II data, see e.g. talk by G. Pearce, at ICHEP'00, 
   Osaka (2000).

\bibitem{Chooz}  CHOOZ collaboration, M. Apollonio et al., Phys.\ Lett.\ 
   B466, 415, (1999), hep-ex/9907037.

\bibitem{K2K} S.H. Ahm et al., Phys. Lett. B511, 178, (2001), hep-ex/0103001.

\bibitem{Sno} Q.R. Ahmad et al. Phys. Rev. Lett. 87, 071307, (2001),
   nucl-ex/0106015.

\bibitem{Fogli}  G.L. Fogli et al., hep-ph/0106247, Phys. Rev. D64,
   0093007 (2001).

\end{thebibliography}
\end{document}